\documentclass{iau}

\usepackage{amsmath}  
\usepackage{graphicx}
\usepackage{multirow}
\usepackage{orcidlink}

\newcommand{\adv}{    {\it Adv. Sp. Res.}} 
\newcommand{\aap}{    {\it A\&A}}
\newcommand{\apj}{    {\it ApJ}}
\newcommand{\apjl}{   {\it ApJ Lett.}}
\newcommand{\an}{     {\it AN}} 
\newcommand{\jgr}{    {\it J. Geophys. Res.}}
\newcommand{\prl}{    {\it Phys.~Rev.~Lett.}}
\newcommand{\solphys}{{\it Solar Phys.}}
\newcommand{\etal}{{\it et al.}}

\newcommand{\kms} {\mathrm{km\,s}^{-1}} 
\newcommand{\vcme}{V_\mathrm{CME}}
\newcommand{\Bep} {B_\mathrm{ep}}
\newcommand{\Bet} {B_\mathrm{et}}
\newcommand{\Bpot}{B_\mathrm{pot}}
\newcommand{\VA}  {V_\mathrm{A}}
\newcommand{\ncr} {n_\mathrm{cr}}
\newcommand{\hcr} {h_\mathrm{cr}}

\begin{document}

\lefttitle{Bernhard Kliem \textit{\etal}}
\righttitle{Magnetic Decay Index Profile and Coronal Mass Ejection Speed}

\jnlPage{1}{7}
\jnlDoiYr{2024}
\doival{10.1017/xxxxx}
\volno{388}
\pubYr{2025}
\journaltitle{Solar and Stellar Coronal Mass Ejections}

\aopheadtitle{Proceedings of the IAU Symposium}
\editors{N. Gopalswamy,  O. Malandraki, A. Vidotto \&  W. Manchester, eds.}

\title{Magnetic Decay Index Profile and Coronal Mass Ejection Speed}

\author{Bernhard Kliem$^{1,2}$\orcidlink{0000-0002-5740-8803}, 
        Georgios Chintzoglou$^3$\orcidlink{0000-0002-1253-8882}, 
        Tibor T{\"o}r{\"o}k$^4$\orcidlink{0000-0003-3843-3242}, and 
        Jie Zhang$^2$\orcidlink{0000-0003-0951-2486}}
\affiliation{$^1$Institute of Physics \& Astronomy, University of Potsdam, 14476 Potsdam, Germany}
\affiliation{$^2$George Mason University, Fairfax, VA, USA}
\affiliation{$^3$Lockheed Martin Solar \& Astrophysics Laboratory, Palo Alto, CA, USA}
\affiliation{$^4$Predictive Science, Inc., San Diego, CA, USA}

\begin{abstract}
We study the relationship between the speed of coronal mass ejections (CMEs) and the height profile of the ambient magnetic field, quantified by its decay index, $n(h)$. Our sample is composed of 15 very fast CMEs ($\vcme\ge1500~\kms$; all halo CMEs) and 22 halo CMEs below this speed limit from Solar Cycle 23. The very fast CMEs yield a high correlation of 0.81 between $\vcme$ and the slope of $n(h)$ in a height range above the onset height of the torus instability if one extremely fast outlier, which closely followed another very fast CME, is excluded. This is consistent with the hypothesis that the torus instability plays a decisive role in CME acceleration. The whole sample yields a weaker correlation, which is still significant if events with a broad torus-stable dip in $n(h)$ are excluded. 
A parametric simulation study of flux-rope eruptions from quadrupolar and two-scale bipolar source regions confirms the decelerating effect of such dips. Very fast, moderate-velocity, and confined eruptions are found.
\end{abstract}


\maketitle

\section{Introduction}\label{s:intro}

The speed of coronal mass ejections (CMEs), $\vcme$, is a major factor in controlling their geoeffectiveness. It is decisive for the formation of CME-driven shocks, which may produce large solar energetic particle events, strongly influences (joint with the frontside field direction and ram pressure) the strength of the geomagnetic perturbation in cases of direct interaction with the magnetosphere, and is correlated with the magnitude of the usually associated flare \citep{srivastava04, gopalswamy08, ZhangJ&al2021}. Most fast CMEs ($\vcme>700\kms$) are fully accelerated in the inner corona, within $h<1R_\odot$ above the photosphere \citep{macqueen83, zhang.j06}. Therefore, the capability to predict the coronal velocity of CMEs is of great practical interest. 

Source-region parameters that were found to correlate with CME speed mostly refer to the photospheric magnetic field and can be divided into \emph{magnitude parameters} such as magnetic flux, area, average field strength, \emph{structural parameters} such as the number of polarity inversion lines (PILs), a measure of complexity \citep{wang.y08}, the effective distance of the main polarities \citep{guo.j06}, and the change of the shear angle at the PIL during the eruption \citep{su07}, and \emph{combined parameters} that scale with both the magnitude of the source region and its structural properties, such as the free magnetic energy and helicity of the source region \citep{liu.y07a, sung09}, PIL length \citep{wang.y08}, and non-neutralized current \citep{Kontogiannis&al2019, liu.y24}. Parameters of the coronal field have received relatively little attention so far, with CMEs originating under pseudostreamers (more open field) being faster on average than those originating under regular streamers \citep{liu.y07} and (probably related to this) higher decay index of the coronal field being associated with faster CMEs \citep{xu.y12, DengM&Welsch2017}. 
CME speeds are strongly correlated with the amount of flux reconnected in the associated flare \citep{qiu05, DengM&Welsch2017, ZhuC&al2020}. However, the cause-effect relationship---whether stronger reconnection causes a faster rise of the CME flux rope or vice versa---is not yet known. 

All associations with source region parameters show scatter, especially when the sample is not restricted to fast CMEs, indicating that $\vcme$ is controlled by a combination of source-region parameters and that non-magnetic forces (aerodynamic drag) may play a significant role for the slower CMEs within the coronagraph's field of view. The highest correlation coefficients, $c\gtrsim0.8$ for fast CMEs, were found between $\vcme$ and PIL length, total non-neutralized current, and reconnected flux \citep{Kontogiannis&al2019, qiu05}. 

\begin{table}[t]                                                        
\centering
\caption{Event sample} 
{\tablefont\begin{tabular}{@{\extracolsep{\fill}}rrrrrrr}
\midrule 
{Event} &    Date     &    Time    &  NOAA  & $\vcme$& $\hcr$ & $\langle n^\prime(h) \rangle_{{\Delta}h}$ \\ 
        &             &    [UT]    &    AR  &[$\kms$]&  [Mm]  &  [$R_\odot^{-1}$] \\ 
\midrule 
  0     & 2000/07/14  &  10:54:07  &  9077  &  1674  &    51  &    5.2 \\
  1     & 2000/09/12  &  11:54:05  &  9163  &  1550  &   177  &    2.4 \\
  2     & 2001/09/24  &  10:30:59  &  9632  &  2402  &    51  &   13.7 \\
  3     & 2002/08/16  &  12:30:05  & 10069  &  1585  &   135  &    3.5 \\
  4     & 2002/09/05  &  16:54:06  & 10102  &  1748  &   147  &    4.0 \\
  5     & 2002/11/09  &  13:31:45  & 10180  &  1838  &    42  &    9.4 \\
  6     & 2003/10/28  &  11:30:05  & 10486  &  2459  &    59  &   10.8 \\
  7     & 2003/10/29  &  20:54:05  & 10486  &  2029  &    55  &   12.4 \\
  8     & 2003/11/18  &  08:50:05  & 10501  &  1660  &    97  &    4.1 \\
  9     & 2004/11/07  &  16:54:05  & 10696  &  1759  &    59  &    8.1 \\
 10     & 2005/01/15  &  06:30:05  & 10720  &  2049  &    46  &    8.4 \\
 11     & 2005/01/15  &  23:06:50  & 10720  &  2861  &    51  &    6.0 \\
 12     & 2005/01/17  &  09:30:05  & 10720  &  2094  &    55  &    5.5 \\
 13     & 2005/05/13  &  17:12:05  & 10759  &  1689  &    76  &    5.7 \\
 14     & 2006/12/13  &  02:54:04  & 10930  &  1774  &    59  &    5.5 \\
 15     & 1997/05/12  &  05:30:05  &  8038  &   464  &    46  &    8.5 \\
 16     & 1997/08/30  &  01:30:35  &  8076  &   371  &   164  &    1.7 \\ 
 17     & 1997/09/28  &  01:08:33  &  --    &   359  &   160  &    2.0 \\
 18     & 1999/06/12  &  21:26:08  &  8569  &   465  &   139  &    2.8 \\
 19     & 2000/07/07  &  10:26:05  &  9070  &   453  &    80  &    9.1 \\
 20     & 2000/09/15  &  21:50:07  &  9165  &   257  &    29  &   29.3 \\
 21     & 2001/03/19  &  05:26:05  &  9380  &   389  &   206  &    0.6 \\
 22     & 2000/02/17  &  21:30:08  &  8872  &   728  &    93  &    4.4 \\
 23     & 2001/03/24  &  20:50:05  &  9390  &   906  &    72  &    4.8 \\
 24     & 2001/09/28  &  08:54:34  &  9636  &   846  &   118  &    0.0 \\
 25     & 2001/11/21  &  14:06:05  &  9704  &   518  &    63  &    1.2 \\
 26     & 2002/04/15  &  03:50:05  &  9906  &   720  &    67  &    2.6 \\
 27     & 2005/05/26  &  15:06:05  & 10767  &   586  &    46  &    4.9 \\
 28     & 2006/04/30  &  09:54:04  & 10876  &   544  &   114  &    4.5 \\
 29     & 2000/02/08  &  09:30:05  &  8858  &  1079  &   244  &    2.7 \\
 30     & 2000/02/12  &  04:31:20  &  8858  &  1107  &   164  &    1.0 \\
 31     & 2000/06/06  &  15:54:05  &  9026  &  1119  &    59  &    1.4 \\
 32     & 2000/11/24  &  05:30:05  &  9236  &  1289  &    55  &   15.2 \\
 33     & 2001/04/09  &  15:54:02  &  9415  &  1192  &   223  &    2.9 \\
 34     & 2001/10/22  &  15:06:05  &  9672  &  1336  &    51  &   -0.9 \\
 35     & 2002/12/19  &  22:06:05  & 10229  &  1092  &    76  &   14.0 \\
 36     & 2005/05/06  &  17:28:31  & 10758  &  1128  &   122  &    2.3 \\
\midrule 
\end{tabular}}
\tabnote{\textit{Notes}: The sample consists of 15 very fast ($\vcme\ge1500~\kms$) and 22 slower halo CMEs from Solar Cycle 23. Time of first appearance in the \textsl{SOHO}/LASCO/C2 coronagraph \citep{brueckner95}, source region (NOAA active region [AR] number), and linear CME speed in the combined LASCO/C2 \& C3 field of view are given. Parameter values maximizing the correlation between $\vcme$ and $\langle n^\prime(h) \rangle_{{\Delta}h}$ for the very fast CMEs, $\ncr=1.5$ and $\delta=1$, are used in the computation of critical height, $\hcr$, and average slope of the decay index, $\langle n^\prime(h) \rangle_{{\Delta}h}$.}
\label{t:sample}
\end{table}

Here, we study the dependence of $\vcme$ on the height profile of the coronal field. Assuming a major role for the torus instability (TI) in the acceleration of CMEs \citep{kliem06}, one can expect that the height profile of the decay index, $n(h)=-\mathrm{d}\log{B_\mathrm{ep}(h)}/\mathrm{d}\log{h}$, above the critical height for TI onset, $h>\hcr$ (determined by $n(h)>n_\mathrm{cr}$), is related to CME acceleration. For a steeper height profile, the external poloidal (``strapping'') field, $\Bep$, decreases faster with height, so that the mismatch between the upward Lorentz self-force of the erupting flux rope and the downward tension force of the strapping field rapidly increases, equivalent to strong CME acceleration and suggesting a correlation between the slope $n^\prime(h\!\ge\!\hcr)$ and $\vcme$. The TI threshold depends on parameters and is usually found to lie in the range $\ncr\approx1\mbox{--}2$, with a canonical value of 3/2. Because $n(h)$ is a structural parameter of the source region, its slope can only control the resulting velocity of a torus-unstable flux rope relative to the Alfv\'en velocity, $\VA$, in the source volume, which limits the correlation with $\vcme$. Nevertheless, $n^\prime(h\!\ge\!\hcr)$ deserves attention as a potentially important parameter in the acceleration of CMEs.

\section{Data analysis}\label{s:data_analysis}

\begin{figure}[t]                                                       
\centering 
\includegraphics[width=.95\linewidth]{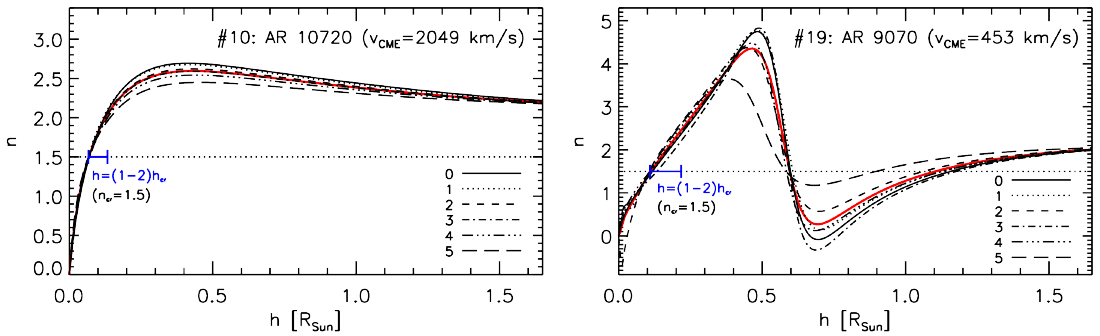}  
\caption{
Decay index height profiles for the CMEs on 2005-01-15, 06:30~UT (left) and 2000-07-07, 10:26~UT (right). Each case has 6 sampling lines, and the average $n(h)$ is plotted in red. The height range $\Delta h$ used to compute $\langle n^\prime(h) \rangle_{{\Delta}h}$ is marked in blue.} 
\label{f:profiles}
\end{figure}

We extract a sample of 15 very fast front-side CMEs ($\vcme\ge1500~\kms$), all halo CMEs, from the complete sample of 57 such CMEs in 1996/06--2007/01 in \citet{wang.y08} by restricting the source location to $\le30^\circ$ from disk center (for reliable magnetogram data) and by requiring that flare ribbon observations are available. We add slower CMEs from the CDAW Halo CME catalog \citep{gopalswamy10a} using the same selection criteria, in a first step limited to 22 additional events with good data of the flare ribbons, to allow careful individual analysis of each event in this sub-sample. 
The events are listed in Table~\ref{t:sample}. 

From the extent of the flare ribbons, we determine the erupting section of the PIL and place 3--7 nearly equally distributed points along that section. The coronal potential field, $\Bpot$, is computed along vertical ``sampling'' lines rooted in these points using a spherical-harmonic expansion (with 248 harmonics for high accuracy) and a source surface, where the field is prescribed to be radial, at $r=2.66~R_\odot$ \citep{schatten69, altschuler69}. For each CME, the synoptic map from the Michelson Doppler Imager \cite[MDI,][]{scherrer95} aboard \textsl{SOHO} is patched with the final pre-eruption MDI magnetogram of the source region. The line-of-sight field component is converted to the radial one, assuming that the photospheric field is approximately radial \citep{wang.y.m92}. 

The decay index along each sampling line is computed using the horizontal component of $\Bpot$ as an approximation of $\Bep$. Averaging over all sampling lines for each source region yields one function $n(h)$ for each CME. Different from many earlier studies, which used a fixed height range in the analysis of $n(h)$, often $h=42\mbox{--}105$~Mm, we use an event-specific height range, $\Delta h$, by determining $\hcr$ for each event individually (Table~\ref{t:sample}) and setting $\Delta h \in [h_0,h_\mathrm{max}]=[1,1+\delta]\hcr$. 
The free parameter $\delta$ is set uniformly across the sample, but a range of values, $0.1\le\delta\le2$, is considered in the analysis, to check the robustness of the results. For the same purpose, TI threshold values in the range $1.3\le\ncr\le1.7$ are considered. The upper edge of the height range is limited by the location of the source surface in the $\Bpot$ model. The height range scales with $\hcr$, hence, with the related main spatial scale of the source region; this improves the correlations found. Finally, the average slope of the decay index profile in the range $\Delta h$, $\langle n^\prime(h) \rangle_{{\Delta}h}$, is computed. Figure~\ref{f:profiles} illustrates $n(h)$ and $\Delta h$ for two representative events and our standard values of $\ncr=1.5$ and $\delta=1$.

\section{Observational results}\label{s:obs_results}

\begin{figure}[t]                                                       
\centering
\includegraphics[width=0.95\linewidth]{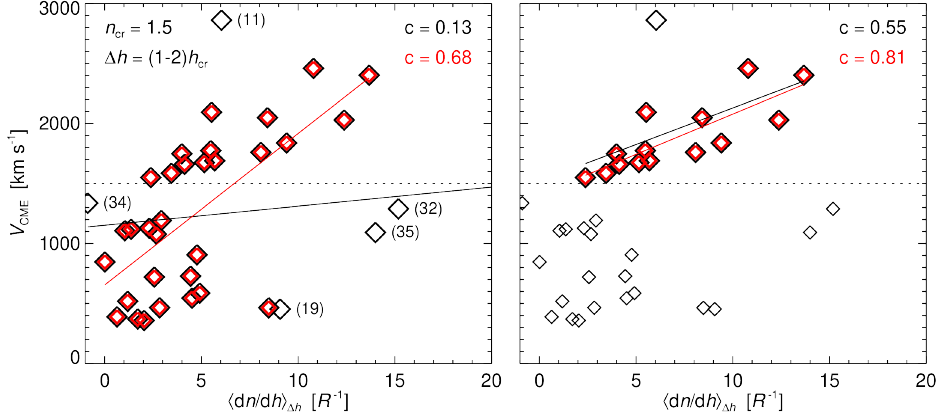}  
\caption{
$\vcme$ vs.\ $\langle n^\prime(h) \rangle_{{\Delta}h}$ for the parameters $\ncr=1.5$ and $\delta=1$ that maximize the correlation coefficients for the very fast CMEs. The whole sample yields a correlation coefficient $c=0.13$ (left panel, black line is the linear fit). Excluding Event~\#11 and all 5 events with a strong dip in $n(h)$ (\#19, 20, 32, 34, 35; open diamonds; see text), raises the correlation to $c=0.68$ (red fit line). Event \#20 lies outside the range shown (see Table~\ref{t:sample}). 
Results for the sub-sample of very fast CMEs are shown in the right panel.} 
\label{f:corr_all} 
\end{figure}

Figure~\ref{f:corr_all} shows that $\langle n^\prime(h) \rangle_{{\Delta}h}$ and $\vcme$ are significantly correlated for the sub-sample of very fast CMEs ($c=0.55$), consistent with the TI playing a significant role in their acceleration. 
Similar to all other parameters studied in the literature, the correlation degrades when the whole range of CME velocities is included. However, the correlations for both samples are considerably improved when two categories of events are excluded. The first of these is a CME (\#11) which is particularly fast, the likely reason being that it followed another fast CME from the same source region (\#10 [Fig.~\ref{f:profiles} left]) within 16.5 hours; the resulting higher acceleration would then not primarily be 
related to the $n(h)$ profile, but rather to a coronal density depletion. The second are cases with a dip in $n(h)$ deep and broad enough to introduce a significant torus-stable height range above the unstable height range between $\hcr$ and the dip (\#19 [Fig.~\ref{f:profiles} right], 20, 32, 34, 35). Flux ropes erupting from the unstable height range just above $\hcr$ are significantly decelerated in such overlying torus-stable height ranges. If they still enter the upper torus-unstable range, a CME results, otherwise the eruption remains confined. As Fig.~\ref{f:profiles} illustrates, the corresponding $n(h)$ profiles tend to be steep between $\hcr$ and the peak value of $n(h)$ below the dip (as explained in Sect.~\ref{s:simulation}), producing the high-slope lower-speed data points in Fig.~\ref{f:corr_all}, which, consequently, deviate strongly from the expected correlation between $\langle n^\prime(h) \rangle_{{\Delta}h}$ and $\vcme$. The resulting correlation for the very fast CMEs, $c=0.81$ with a confidence interval of [0.69,0.89] at the $1\sigma$ (68\%) confidence level, belongs to the highest obtained so far for any source-region parameter in spite of the fact that $\langle n^\prime(h) \rangle_{{\Delta}h}$ is only a structural parameter not covering the spread in source-region Alfv\'en speed. The correlation coefficient changes (decreases) by no more than 10\% in the range $\ncr=1.3\mbox{--}1.7$ if $\delta=1$ is used and by 25\% in the whole range of the free parameters considered. The correlation for the whole sample is also significantly improved to $c=0.68$ (with a confidence interval of [0.57,0.77]) by excluding the events with dipped $n(h)$ and Event~\#11.

\section{Simulation study}\label{s:simulation}

\begin{figure}[t]                                                       
\centering
\includegraphics[width=0.95\linewidth]{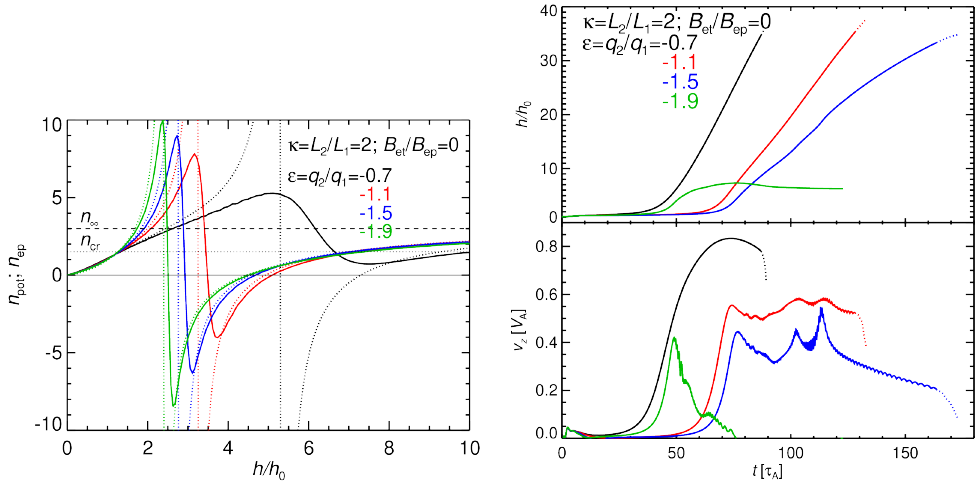}  
\caption{Decay index profiles $n(h)$ (left) and rise profiles of TD flux ropes, $h(t)/h_0$ and $v_z(t)/\VA$, (right) in a line quadrupole with $\Bet=0$ for a range of source strength ratios $\epsilon=q_2/q_1$. Dotted profiles $n(h)$ are the exact profiles computed from $\Bep$, which have a pole at the position of the null (X-) line, and the solid profiles are computed from the corresponding potential field. The rise profiles are dotted during the period when the flux rope apex closely approaches the closed upper boundary.} 
\label{f:n&cfl_all_quadrupole} 
\end{figure}

In order to quantify the effect of dipped $n(h)$ profiles on $\vcme$ and to establish criteria for their exclusion from our correlation analysis, we have started a parametric simulation study of flux rope eruptions. Dipped $n(h)$ profiles occur in multipolar regions with a magnetic null line or point, in particular in quadrupolar regions \citep{torok07}, and in two-scale bipolar regions if the scales of the photospheric flux distribution are sufficiently different \citep{Kliem&al2021}. Both configurations can be modeled through a simple extension of the Titov-D\'emoulin force-free flux rope equilibrium \cite[][henceforth TD99]{Titov&Demoulin1999} by replacing the sub-photospheric bipole, which is the source of $\Bep$, with a pair of bipoles, both placed symmetrically to the plane of the toroidal flux rope and on the axis of rotational symmetry. Denoting the magnetic charges as $q_1$ and $q_2$ and the half-spacings of the bipoles as $L_1$ and $L_2$, a line quadrupole results for $q_1q_2<0$ ($L_2>L_1$; see Fig.~5 in \citealt{Kliem&al2014} for an illustration) and a two-scale bipolar region results for $q_1q_2>0$, $L_2\gg L_1$. These configurations are used as the initial condition of zero-beta ideal MHD simulations in large boxes of size $40^3$ constructed with a stretched grid of highest resolution 0.04 in the area of the initial flux rope, whose apex height is normalized to $h_0=1$ (see, e.g., \citealt{Hassanin&al2022} for the numerical details). The ratios $\epsilon=q_2/q_1$ and $\kappa=L_2/L_1$ are used as independent parameters and $q_1$, $q_2$ are determined by the equilibrium condition (TD99) for given geometry ($L_1$, $L_2$), which is chosen such that the equilibrium is very close to marginal stability. A short, transient phase of initial numerical relaxation of the approximate analytical equilibrium causes the flux rope to expand slightly, whereby it enters the unstable domain, so that the TI develops spontaneously. 

Figure~\ref{f:n&cfl_all_quadrupole} shows the decay index and rise profiles for a TD flux rope embedded in a line quadrupole with zero external toroidal (shear) field, $\Bet=0$. The X-line above the flux rope thus turns into a null line, where $n(h)$ has a pole, which introduces a dip above the pole and leads to a high slope $n^\prime(h)$ below the pole. When $n(h)$ is approximated by using the horizontal components of $\Bpot$ instead of $\Bep$, as in our observational study, the pole turns into an inflection point with a dip similar in shape to our dipped observational $n(h)$ profiles (Fig.~\ref{f:profiles}). $\Bpot$ is computed in the box from the initial TD field values at the boundaries. The flux in the footprints of the flux rope acts as the source of a guide field component in the potential field along the null line of $\Bep$, turning the pole into an inflection point. As $\epsilon$ increases, corresponding to increasing exterior flux, 
the dip in $n(h)$ turns deeper and wider, and shifts to smaller heights. The corresponding rise profiles show an increasing deceleration of the erupting flux rope until a confined eruption results for the highest $\epsilon$ chosen; they depend only weakly upon $\kappa$. 

\begin{figure}[t]                                                       
\centering
\includegraphics[width=0.95\linewidth]{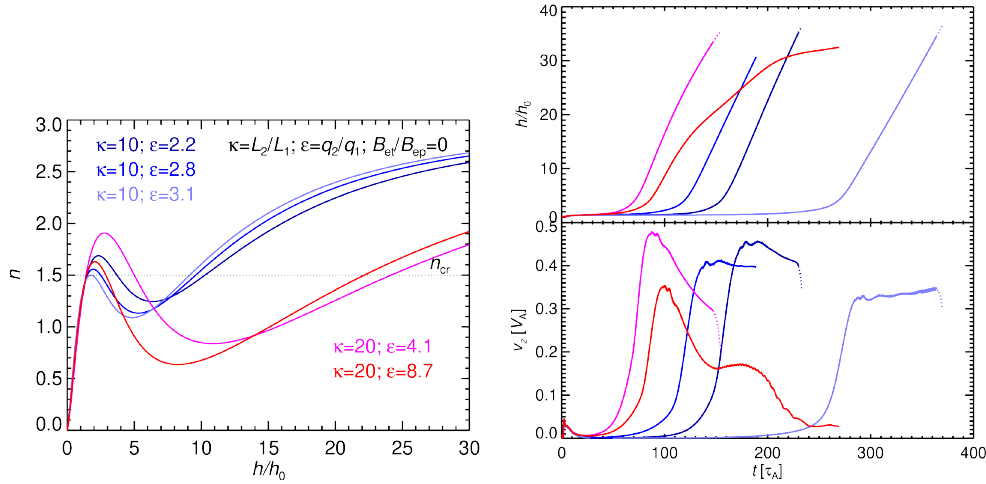}  
\caption{Decay index and rise profiles for the two-scale bipole with $\Bet=0$ for a range of source strength ratios $\epsilon=q_2/q_1$ and two values of the scale ratio $\kappa=L_2/L_1$, displayed as in Fig.~\ref{f:n&cfl_all_quadrupole}. The very different evolution times until considerable acceleration commences result from small differences of the initial equilibria from the exact marginal stability configurations.} 
\label{f:n&cfl_all_bipole} 
\end{figure}

Figure~\ref{f:n&cfl_all_bipole} shows the results for the two-scale bipole. Again, a broader and deeper dip in $n(h)$ increasingly decelerates the torus-unstable flux rope until a confined eruption results for the maximum values of $\epsilon$ and $\kappa$ shown. The minimum scale ratio for the effect to occur is found to be $\kappa\sim10$. Both simulation series confirm our rationale for excluding cases with deeply dipped $n(h)$ from the correlation analysis in Sect.~\ref{s:obs_results}. 

\section{Discussion}\label{s:discussion}

By restricting the study to halo CMEs with good data of flare ribbons, the sub-sample of slower CMEs ($\vcme<1500~\kms$) is strongly reduced in size compared to the whole population of slower CMEs. However, to provide reliable support to the forecast of very fast CMEs, our ultimate goal, the present study will be extended to include all CMEs with usable flare ribbon data from a representative period. With such a larger sample, it remains to be seen whether the exclusion of high-slope, lower-speed events based on the properties of a dip in $n(h)$ can be nearly as complete as with the sample used here. 

The parametric simulation study of eruptions from source regions with dipped $n(h)$ profiles will be extended to quantify the relationship between the properties of $n(h)$ (torus-stable dip and underlying torus-unstable peak) and the resulting velocity of erupting flux ropes (normalized by $\VA$). This will be related to the ratio of core-region flux forming the equilibrium and overlying exterior flux, and ultimately to the flux distribution in the photosphere. 

For applicability in forecast schemes, a method to estimate the potentially erupting section of the PIL must be included. The computation of the coronal currents and their neutralization degree might indicate the section most clearly, but depends on a nonlinear force-free field extrapolation, which is computationally demanding and does not yield unique solutions. Selecting the strong-gradient section of the PIL \citep{Falconer&al2003, Schrijver2007, Dhakal&ZhangJ2024} is a promising option. 

The correlation between $\langle n^\prime(h) \rangle_{{\Delta}h}$ and $\vcme$ might improve further if oblique propagation directions, which occur for many CMEs, are considered. The expected propagation direction can be estimated from the asymmetry of flux in the environment of the erupting PIL section.

\section{Conclusions}\label{s:conclusions}

From the presented stage of our ongoing study, we conclude the following. 
\vspace{-\baselineskip} 
\begin{enumerate} 
\item $\langle n^\prime(h) \rangle_{{\Delta}h}$ shows a very high correlation with $\vcme$ for very fast CMEs ($\vcme\ge1500~\kms$) with $c=0.81$ in Cycle 23, among the highest correlations found so far, although $\langle n^\prime(h) \rangle_{{\Delta}h}$ is only a structural parameter not covering the spread in $\VA$. 
\item This is consistent with the hypothesis that the TI plays a decisive role in the coronal acceleration of CMEs. 
\item A significant torus-stable dip in $n(h)$, reflecting a significant amount of exterior flux passing over the current-carrying core flux, indicates a deceleration of flux ropes erupting from the core flux, including the possibility of confinement. 
\end{enumerate} 
\textbf{Competing interests.} The authors declare none. \\ 
\textbf{Acknowledgments.} We acknowledge very helpful comments by an anonymous referee and support from the DFG and from NASA through grant No.\ 80NSSC20K1274.

\end{document}